\documentclass[RNAAS]{aastex62}

\begin{document}

\title{P/2017 S5: another active asteroid associated with the Theobalda family}

\correspondingauthor{Bojan Novakovi\'{c}}
\email{bojan@math.rs}

\author[0000-0001-6349-6881]{Bojan Novakovi\'{c}}
\affiliation{Department of Astronomy, Faculty of Mathematics, University of Belgrade \\
Studentski trg 16, 11000 Belgrade, Serbia \\}



\section{} 

Active asteroids are a relatively new class of objects in the asteroid belt. These intriguing objects have both, the orbital characteristics of asteroids and the physical characteristics of comets \citep{Jewitt2015aste.book}. Main belt comets (MBCs) are a subgroup of active asteroids for which it is believed that observed activity is driven by the sublimation of volatile ices \citep{Hsieh2006Sci,Snodgrass2017A&ARv}. Recent work by \citet{Hsieh2018AJ} has demonstrated that all MBCs associated to collisional families belong to families with primitive taxonomic classifications.

In September 2017 a new active asteroid has been discovered, namely P/2017 S5 (hereafter S5) \citep{Sato2017CBET}. Aiming to better constrain the orbital classification of S5 and to cast around for an associated family, we first compute its proper orbital elements \citep{Zoran2017SerAJ}. The proper elements are computed starting from the nominal osculating elements and following procedure described in \citet{KM2000}. The current orbit solution provided by \href{https://ssd.jpl.nasa.gov/sbdb.cgi}{JPL} is based on 411 observations distributed over an arc of $139^\circ$. The orbital parameters and their corresponding $1-\sigma$ uncertainties at epoch 2458050.5 are: semi-major axis, $a = 3.1709964417 \pm 3.5836 \times 10^{-5}$~au, eccentricity, $e = 0.3131090886 \pm 8.1092 \times 10^{-6}$, inclination, $i = 11\fdg 849111891 \pm 0\fdg 00013861$, longitude of the ascending node, $\Omega$ = $252\fdg 392562 \pm 0\fdg 000552$, argument of perihelion, $\omega$ = $99\fdg 917748 \pm 0\fdg 003879$, and mean anomaly, $M = 15\fdg 36782 \pm 0\fdg 00163$.
The calculated proper orbital elements are: semi-major axis, $a_p = 3.185133$~au, eccentricity, $e_p = 0.255887$ and sine of inclination $\sin(i_p) = 0.247459$. These values are consistent with orbital characteristics of main belt asteroids.

In the next step we search for a possible dynamical family around S5.
Using the catalog of proper elements for main belt asteroids available at \href{http://asteroids.matf.bg.ac.rs/fam/hcm.php}{Asteroid Families Portal} and computed proper
elements for S5, we applied the Hierarchical Clustering Method (HCM) proposed by \citet{Zappala1990} with S5 as a central body \citep[see][for details on this approach]{Viktor2017MNRAS}. The results suggest that S5 is a member of Theobalda family \citep{Milani2014Icar}, estimated to be about 7~Myr old \citep{Nov2010MNRAS}. 
S5 joins the family at cut-off distance of 20~m/s, that is well below the nominal cut-off value of 55~m/s provided at \href{http://asteroids.matf.bg.ac.rs/fam/hcm.php}{Asteroid Families Portal}. This suggests a firm link between the active asteroid and the family.

\begin{figure}[h!]
\begin{center}
\includegraphics[scale=0.4,angle=-90]{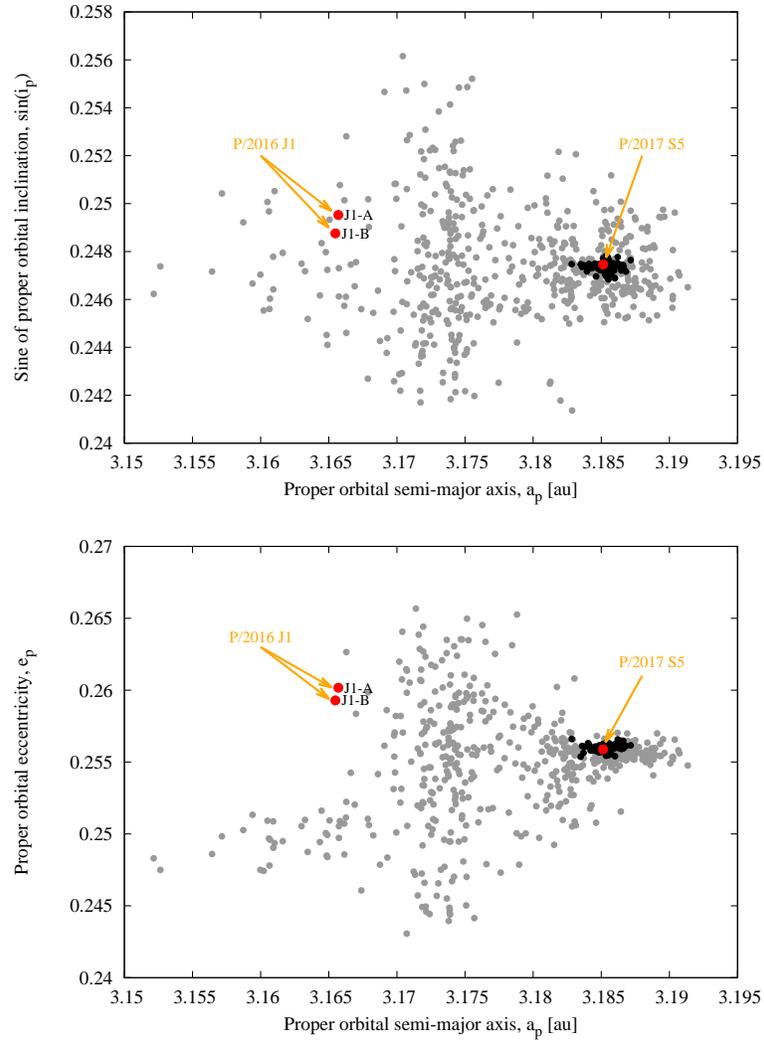}
\caption{The orbital distribution of the Theobalda family members (gray dots) and locations of two main-belt comets (red dots) within the family. The black dots show the proper elements computed for the orbital clones of S5.
The current membership of the Theobalda family has been obtained applying the HCM online at \href{http://asteroids.matf.bg.ac.rs/fam/hcm.php}{Asteroid Families Portal}.\label{fig:1}}
\end{center}
\end{figure}

The current uncertainties of osculating orbital parameters of S5 are however large enough to cast some doubt on the established dynamical connection with the Theobalda family.
In order to test robustness of our findings, we generate 100 orbital clones distributed based on S5's orbital uncertainties \citep[see][for more details on our clone-generation procedure]{Moreno2017ApJ} and compute the proper elements for all of them. Then we apply the HCM using the proper elements of orbital clones, on a case by case basis. We found that each of 100 cloned orbits is dynamically associated to the Theobalda family (Figure~\ref{fig:1}), proving that orbital uncertainties do not affect association of S5 to this family.

Intriguingly, S5 is not the first known active asteroid belonging to this family.
A double component main-belt comet P/2016 J1 \citep{Moreno2017ApJ} is also a member of the Theobalda family \citep{Hsieh2018AJ}. This makes the Theobalda family the third group
known to contain at least two active asteroids, along with the Themis and Lixiaohua families.

The activity driver in case of S5 is still unsure, but based on the object's location
in orbital space and its dynamical association to the primitive Theobalda family, its activity may be driven by the sublimation of water ices. Moreover, its brightness profile is consistent with a model where dust is launched from the nucleus isotropically at a constant speed, in agreement with an assumption that the activity is driven be the sublimation of volatiles \citep{Borysenko2018Icar}. If the above hypothesis is confirmed by future data, this would further strengthen the link between MBCs and primitive asteroid families.

\acknowledgments

This work has been supported by the Ministry of Education, Science and Technological Development of the Republic of Serbia, under the Project 176011.

\end{document}